\documentclass[aps,prl,twocolumn,amsmath,amssymb,showpacs]{revtex4-1}
\usepackage[ansinew]{inputenc}
\usepackage{graphicx}
\usepackage{textcomp}
\usepackage{hyperref}

\newcommand{\ket}[1]{|#1\rangle}

\begin{document}

\title{Ground-State Cooling of a Single Atom at the Center of an Optical Cavity}
\author{Andreas~Reiserer}
\author{Christian~N\"olleke}
\author{Stephan~Ritter}
\email{stephan.ritter@mpq.mpg.de}
\author{Gerhard~Rempe}

\affiliation{Max-Planck-Institut f\"ur Quantenoptik, Hans-Kopfermann-Strasse 1, 85748 Garching, Germany}

\begin{abstract}
A single neutral atom is trapped in a three-dimensional optical lattice at the center of a high-finesse optical resonator. Using fluorescence imaging and a shiftable standing-wave trap, the atom is deterministically loaded into the maximum of the intracavity field where the atom-cavity coupling is strong. After 5\,ms of Raman sideband cooling, the three-dimensional motional ground state is populated with a probability of $(89\pm2)\,\%$. Our system is the first to simultaneously achieve quantum control over all degrees of freedom of a single atom: its position and momentum, its internal state, and its coupling to light.
\end{abstract}

\pacs{37.10.De, 37.10.Jk, 37.30.+i, 42.50.Dv, 42.50.Pq}

\maketitle

The investigation and utilization of genuine quantum-mechanical systems require full control over all relevant degrees of freedom. Important for applications is the possibility to initialize the system in a well-defined quantum state and the capability to strongly couple it to other quantum systems. A paradigm system in this context is a single atom coupled to a single photon in a high-finesse optical resonator. In theory, the ideal cavity quantum electrodynamics (CQED) situation assumes a pointlike atom at a fixed position. In practice, however, the motion of the atom deteriorates any localization. In fact, despite long-lasting efforts \cite{ye_trapping_1999, pinkse_trapping_2000} with Doppler and Sisyphus cooling \cite{mckeever_state-insensitive_2003}, cavity \cite{maunz_cavity_2004, nussmann_vacuum-stimulated_2005}, feedback \cite{koch_feedback_2010}, electromagnetically induced transparency \cite{kampschulte_eit-control_2012}, and even one-dimensional (1D) Raman sideband cooling \cite{boozer_cooling_2006}, no single-atom CQED experiment has so far obtained full control over the motional degrees of freedom: position and momentum.

The residual motion of the atom leads to unpredictable fluctuations, which are deleterious to nonclassical phenomena \cite{leach_cavity_2004, rempe_optical_1991} and which limit the photon emission and absorption efficiencies and fidelities in coherent quantum networks \cite{ritter_elementary_2012, nolleke_efficient_2013}. Full control over the position and momentum of single atoms is therefore a long-standing goal in CQED \cite{vernooy_well_1997}. It would not only improve existing experiments but is also of utmost importance for future research. For example, it is an ideal starting point for the cavity-based generation of nonclassical states of motion \cite{zeng_quantum_1994}, the transfer of quantum states between atomic motion and light \cite{parkins_quantum_1999}, and the observation of numerous optomechanical effects \cite{kippenberg_cavity_2008} with single phonons and single photons \cite{rabl_photon_2011, nunnenkamp_single-photon_2011}.

\begin{figure}[b]
\includegraphics[width=1.0\columnwidth]{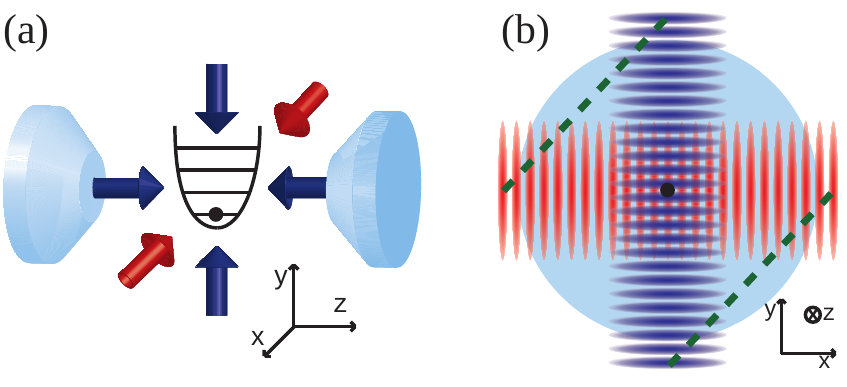}
\caption{\label{fig:setup}
(color online). Sketch of the setup geometry. A single atom (black dot) is trapped at the center of an optical cavity [blue cones in (a), blue disk in (b)] in a 3D optical lattice that is formed by one red-detuned (along $\hat{\text{x}}$) and two blue-detuned (along $\hat{\text{y}}$ and $\hat{\text{z}}$) retroreflected laser beams [arrows in (a), standing-wave structure in (b)]. One Raman beam impinges along the cavity axis ($\hat{\text{z}}$), and two counterpropagating Raman beams impinge orthogonally from the side [area between dashed green lines in (b)], forming an angle of $45^\circ$ with respect to the $\hat{\text{x}}$ and $\hat{\text{y}}$ axes.
}
\end{figure}

In this Letter, we present full control over all degrees of freedom of a single atom strongly coupled to an optical resonator. First, the atom is deterministically localized at a maximum of the cavity field. Second, it is cooled to the three-dimensional (3D) motional ground state. We emphasize that the former is indispensable to reproducibly achieving a constant coupling to the cavity field, while the latter guarantees a constant light shift for an atom in an optical dipole trap and therefore a constant atomic transition frequency.

We realize the ideal CQED situation by trapping a single atom in a three-dimensional optical lattice with the resonator as one of the lattice axes (Fig.\,\ref{fig:setup}). By shifting the standing-wave potential formed by one of the two other lattice beams \cite{nussmann_submicron_2005}, we deterministically localize the atom at the center of the cavity mode. We use high intensities along all three lattice directions to obtain trap frequencies of a few hundred kHz, large compared to the single-photon recoil frequency of about 4 kHz. The atom is then tightly confined (spatial extent of the ground-state wave function $\lesssim 15$\,nm) to the Lamb-Dicke regime (Lamb-Dicke parameter $\eta \lesssim 0.1$) along all three axes \cite{leibfried_quantum_2003} such that a spontaneous emission event most likely does not change the motional state of the atom. This allows us to implement Raman sideband cooling to the 3D ground state in a similar way as has been demonstrated in free space with ensembles of atoms \cite{kerman_beyond_2000, han_3d_2000, grimm_optical_2000}, single ions in Paul traps \cite{diedrich_laser_1989, monroe_resolved-sideband_1995, leibfried_quantum_2003}, and very recently with single atoms in optical tweezers \cite{kaufman_cooling_2012, thompson_coherence_2013}.

The experiment starts with the preparation of a cloud of $^{87}$Rb atoms in a magneto-optical trap. A running-wave dipole trap is then used to transfer the atoms to the optical resonator, where they are loaded into a standing-wave trap (along $\hat{\text{x}}$) at 1064\,nm \cite{nussmann_submicron_2005}. We then apply counterpropagating cooling light perpendicular to the cavity axis ($\hat{\text{z}}$) and at $45^\circ$ with respect to $\hat{\text{x}}$ and $\hat{\text{y}}$. The cooling light is 30\,MHz red-detuned with respect to the $F=2\leftrightarrow F'=3$ transition of the D$_2$ line and has orthogonal linear polarizations. This leads to intracavity Sisyphus cooling \cite{nussmann_vacuum-stimulated_2005} of the atom in all three dimensions. We collect scattered light with a high numerical aperture objective and thereby image the atoms. This allows us to determine their number and position using an algorithm that evaluates the recorded intensity pattern. The loading procedure is repeated until a single atom is detected in the images \cite{bochmann_lossless_2010}. In that case, the standing-wave pattern is shifted along the beam ($\hat{\text{x}}$ axis) such that the atom is transferred to the center of the cavity mode \cite{nussmann_submicron_2005} where the lattice beams tightly confine the atom in 3D.

The two additional lattice beams have a wavelength of around 771\,nm, blue-detuned from the D$_1$ and D$_2$ line. Orthogonal polarization and a sufficient frequency difference prevent any effect of cross interference on the atoms. One of the lattice beams impinges along the $\hat{\text{y}}$ axis through the high numerical aperture objective. The other lattice beam drives a TEM$_{00}$ mode of the cavity, detuned by an odd number of free spectral ranges from the atomic transition at 780\,nm. Therefore, an antinode of the cavity mode coincides with a node of the standing-wave trap light at the center of the cavity, such that the atoms are trapped where the coupling to the resonator is strongest. The coupling strength at a potential minimum changes along the cavity axis due to the different wavelengths of the trapping and cavity field. At a distance of 16.2\,\textmu m from the cavity center, a node of the cavity field coincides with a node of the trap light, such that atoms trapped at this position hardly couple to the cavity.

\begin{figure}
\includegraphics[width=1.0\columnwidth]{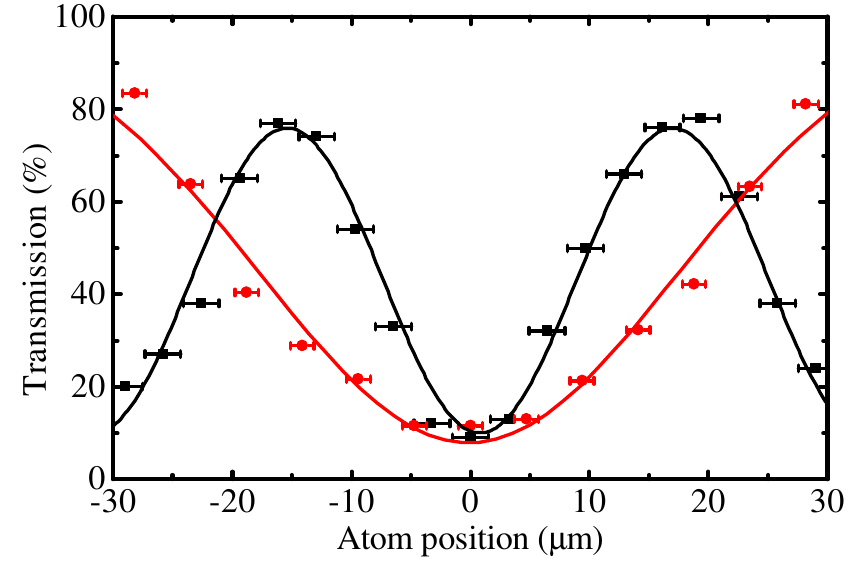}
\caption{\label{fig:positioning}
(color online). Transmission through the cavity when the position of a single, coupled atom is scanned along the cavity ($\hat{\text{z}}$) axis (black squares) and along the $\hat{\text{x}}$ axis (red dots). The transmission is strongly suppressed when the atom is located at a maximum of the intracavity field. The solid Gaussian (red) and sine (black) fit curves are guides for the eye. Note that in this measurement, the atom is not cooled to the ground state.
}
\end{figure}

To demonstrate experimental control of the coupling strength, we load atoms at different positions by shifting the red-detuned dipole trap along the $\hat{\text{z}}$ axis with a piezomirror. Using a $\sigma^+$ polarized laser beam, we optically pump the atom to the $\ket{F,m_F}=\ket{2,2}$ state, where $F$ denotes the hyperfine state and $m_F$ its projection along the quantization axis ($\hat{\text{z}}$). In this state, the atom is coupled to the cavity, which is resonant with the Stark-shifted $\ket{2,2}\leftrightarrow \ket{3,3}$ transition, which has the largest dipole matrix element. Subsequently, the transmission of a weak probe laser resonant with the empty cavity is measured. With our cavity parameters $(g_0, \kappa, \gamma)/2\pi = (8, 3, 3)$\,MHz, the presence of a single, coupled atom leads to a normal-mode splitting  and thus reduces the transmission. Here, $g_0$ denotes the theoretical atom–cavity coupling on the $\ket{2,2}\leftrightarrow \ket{3,3}$ transition for our cavity parameters (radius of curvature 5\,cm, separation 485\,\textmu m), $\kappa$ the cavity field decay rate (finesse $6\times10^4$, free spectral range 309 GHz), and $\gamma$ the atomic polarization decay rate. The beating between the sinusoidal variation of the effective coupling strength $g$ and the standing-wave trap along the cavity axis can be seen in Fig.\,\ref{fig:positioning} (black squares, Stark shift 50\,MHz), where the transmission on resonance is shown as a function of the atomic position along the cavity ($\hat{\text{z}}$) axis \cite{colombe_strong_2007, guthohrlein_single_2001, mundt_coupling_2002}. We observe a sinusoidal modulation of the transmission. The deviation from the ideally expected oscillation with the same period but steeper slopes is caused by a position-dependent optical pumping efficiency and temperature, which leads to averaging effects in coupling strength and Stark shift. A shift of the atom along $\hat{\text{x}}$ (red dots in Fig.\,\ref{fig:positioning}, Stark shift 100\,MHz) and $\hat{\text{y}}$ gives a Gaussian dependence as expected from the radial profile of the cavity mode. Because of the loading procedure, the initial distribution of the atoms in the lattice is determined by their initial temperature and the beam waist of the red-detuned dipole trap. We determine the width of this distribution from the fluorescence images. This gives the error bars in Fig.\,\ref{fig:positioning}. On the length scale of the positioning error, the transmission is nearly constant. We can thus deterministically localize a single atom at the maximum of the resonator field, where the atom-cavity coupling is strongest.

\begin{figure}
\includegraphics[width=1.0\columnwidth]{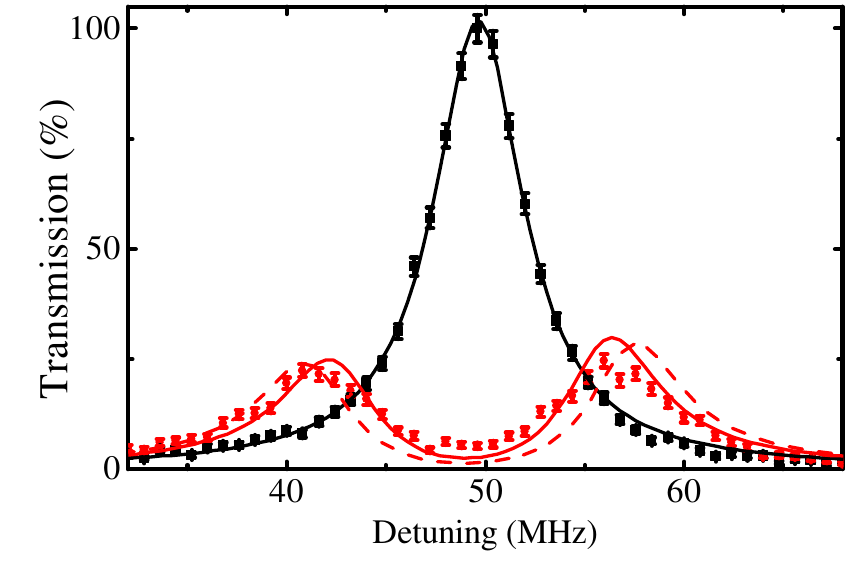}
\caption{\label{fig:normalModes}
(color online). Normal-mode spectroscopy of the atom-cavity system with the atom trapped in the 3D optical lattice. The transmission of the cavity is a Lorentzian curve when the atom is not coupled (black squares and black fit curve), while a resonant atom leads to a normal-mode splitting (red dots and solid red fit curve). The slight asymmetry is caused by a small residual detuning between atom and cavity. The error bars are statistical. The dashed curve shows the spectrum expected for $g_0/2\pi=8$\,MHz, the value calculated from our cavity parameters.
}
\end{figure}

The absolute strength of this coupling is determined by recording the normal-mode spectrum of the atom-cavity system \cite{thompson_observation_1992}. For this purpose, we scan the frequency of a weak probe laser while keeping the frequency of the cavity fixed. To record the spectrum of the empty cavity, we first pump the atom to $F=1$ such that it is not coupled to the resonator. Thus, the transmission is a Lorentzian curve with a full width at half maximum of 5.5\,MHz (Fig.\,\ref{fig:normalModes}, black squares). When the atom is prepared in the $\ket{2,2}$ state, we observe a normal-mode splitting (Fig.\,\ref{fig:normalModes}, red dots). The separation of the two peaks is twice the atom-cavity coupling constant $g$. To determine this value, we fit the normal modes with a theory curve (solid red line) with $g$ and the atomic detuning as the only free parameters. From this fit, we find $g/2\pi=(6.7\pm 0.1)$\,MHz, close to the theoretical value of $g_0/2\pi=8$\,MHz (dashed red line). This again proves that we are able to accurately localize the atom at the center of the cavity field and that the system is in the single-atom strong coupling regime of CQED.

After demonstrating the good control achieved over the position of the atom, we now turn to its motion. In order to measure the temperature, we use Raman sideband spectroscopy. For an atom at low temperature, the sinusoidal lattice can be approximated by a harmonic potential, leading to a quantization of the atomic vibrational energy $\text{E}_\text{\{x,y,z\}}=\left(n_\text{\{x,y,z\}}+\frac{1}{2}\right)h \nu_\text{\{x,y,z\}}$ for each lattice axis \cite{grimm_optical_2000}. Here, $n_\text{\{x,y,z\}}$ is an integer and $\nu_\text{\{x,y,z\}}$ denotes the trap frequency that depends on the intensity and wavelength of the lattice light along the  $\{\hat{\text{x}},\hat{\text{y}},\hat{\text{z}}\}$ direction. We drive transitions between the different motional states using Raman beams. One of the beams is polarized orthogonally to the cavity axis and impinges at an angle of $45^\circ$ with the $\hat{\text{x}}$ and $\hat{\text{y}}$ axis (Fig.\,\ref{fig:setup}). Another beam, polarized along the cavity axis, is counterpropagating to the first one, and a third, also linearly polarized beam is applied along the cavity axis. The latter two are detuned by the hyperfine splitting of 6.8\,GHz from the first, while all of them are red-detuned by 0.3\,THz from the D$_1$ line at 795\,nm. Because of this large detuning, the Raman beams lead to an effective coupling of the two hyperfine ground states without populating the excited state. The linewidth of this coupling can be much smaller than the natural linewidth of the D$_1$ transition. Thus, the sidebands can be spectroscopically resolved and addressed individually when the intensity of the lasers is low enough and the duration of the Raman pulse is long enough.

\begin{figure}[b]
\includegraphics[width=1.0\columnwidth]{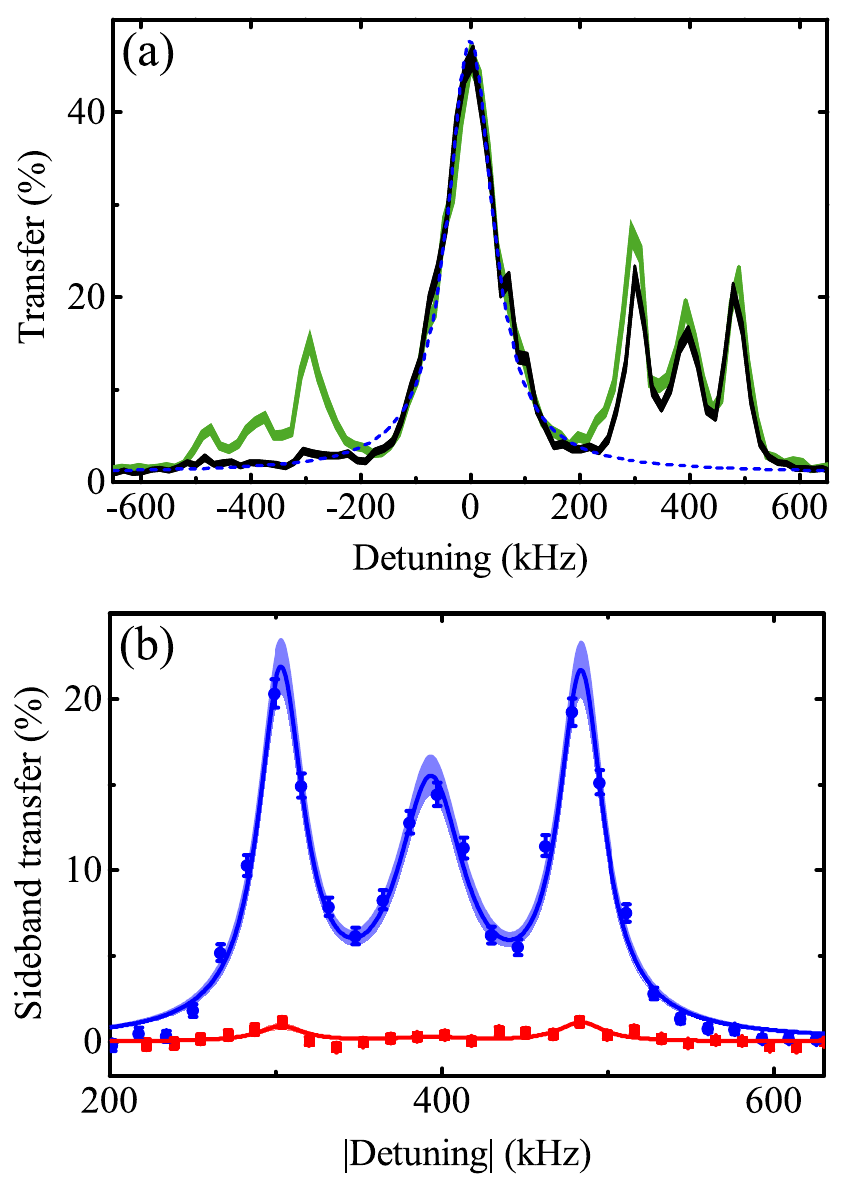}
\caption{\label{fig:SidebandCooling}
(color online). (a) Sideband spectrum after intracavity Sisyphus [green (grey)] and after sideband cooling (black). The statistical standard error of the data is given by the thickness of the lines. The three peaks at positive detunings correspond to a transition on the blue sideband for each axis of the 3D lattice potential (right to left: $\hat x$, $\hat y$, and $\hat z$ axis). The carrier peak at the center (dashed blue Lorentzian fit curve) is saturated. Transitions on the red sideband (negative detunings) are still observed after Sisyphus cooling [green (grey) line] but nearly vanish after 5\,ms of sideband cooling (black line). (b) Transfer probability on the red (red squares) and blue (blue dots) sideband after Raman sideband cooling. The solid curves are numerical fits of the sum of three Lorentzian curves, with the shaded areas indicating the 66\,\% confidence interval. The atomic temperature after sideband cooling is determined from these fits.}
\end{figure}

To obtain a sideband spectrum, the atom is optically pumped to the $F=1$ hyperfine state. Subsequently, we apply the Raman lasers for 200\,\textmu s. In order to measure the population transfer to $F=2$, we perform cavity-based hyperfine-state detection \cite{bochmann_lossless_2010, gehr_cavity-based_2010}. We can thus determine within 30\,\textmu s whether the atomic population has been transferred to $F=2$. The transfer probability after the previously described intracavity Sisyphus cooling as a function of the detuning between the Raman beams is shown in the green (grey) curve of Fig.\,\ref{fig:SidebandCooling}(a), where zero detuning means a frequency difference that corresponds to the hyperfine transition frequency. The large peak at the center of the spectrum is the saturated carrier transition. At negative detunings, the red sidebands can be seen, corresponding to transitions that lower the vibrational state of the atom by one quantum. The three peaks at positive detunings correspond to the blue sideband for each of the three lattice axes: the red-detuned dipole trap along $\hat{\text{x}}$ at 0.5\,MHz and the blue-detuned traps along $\hat{\text{y}}$ and $\hat{\text{z}}$ at 0.4 and 0.3\,MHz, respectively. The peaks can be identified unambiguously by successively changing the intensity of one of the lattice beams and then recording the sideband spectrum (not shown). The central sideband peak ($\hat{\text{y}}$ axis) is lower and broader than the other two in the depicted long-term measurement. On shorter time scales, we observe three peaks of the same height with fluctuating position of the central peak. This is caused by long-term drifts in beam pointing along the $\hat{\text{y}}$ axis, where we use a lattice beam with a much tighter focus due to the limited laser power available. The probability of a change in the vibrational state by $+1$ ($P_\mathrm{blue}$) or by $-1$ ($P_\mathrm{red}$) and thus the height of the peaks in the spectrum is determined by the projection of the $k$-vector difference of the involved Raman beams onto the trap axis and by the population of the different vibrational states $n$ along the respective axis. The ratio between the red and blue sideband peaks gives an upper bound for the mean vibrational state $\bar{n}$ \cite{diedrich_laser_1989, monroe_resolved-sideband_1995, leibfried_quantum_2003, slodicka_interferometric_2012, boozer_cooling_2006, kaufman_cooling_2012,thompson_coherence_2013}:
\begin{equation}
\bar{n}=\frac{P_\mathrm{red}}{ P_\mathrm{blue}-P_\mathrm{red}}.
\end{equation}
Applying this equation to a fit of the green (grey) curve in Fig.\,\ref{fig:SidebandCooling}(a) gives $\bar n_\text{\{x,y,z\}}=\{0.19(5),0.4(1),1.0(2)\}$. This demonstrates that the intracavity Sisyphus cooling we use already leads to temperatures well below the Doppler limit \cite{nussmann_vacuum-stimulated_2005} ($\bar{n}_\mathrm{D} \approx 6-10$ for our trap frequencies).

To further reduce the atomic temperature, we use pulsed Raman sideband cooling. For this purpose, we prepare the atom in $F=1$ and apply the Raman beams for 5\,ms with frequency components that drive transitions on all three red sidebands. During this interval, a $\approx 10$\,ns long repump pulse is applied on the $F=2\leftrightarrow F'=1$ transition every 200\,ns in order to bring any transferred population back to $F=1$, where the cooling cycle can start again. To determine the effect of the sideband cooling, we perform the following measurement cycle: We apply a 4.4\,ms long interval of intracavity Sisyphus cooling on the closed transition. We then record the transfer probability at a certain Raman detuning, apply sideband cooling and again measure the transfer probability at the same detuning. We repeat this measurement sequence at different frequencies to record a spectrum immediately before [green (grey) in Fig.\,\ref{fig:SidebandCooling}(a)] and after (black) sideband cooling. The red sidebands vanish almost completely, which indicates that the atom is cooled close to the ground state. To determine the mean occupation number $\bar n$, we fit a Lorentzian to the carrier [blue dashed line in Fig.\,\ref{fig:SidebandCooling}(a)] and subtract it from the data [Fig.\,\ref{fig:SidebandCooling}(b)]. We fit the sum of three Lorentzian curves to the blue sidebands (blue curve) to determine the width and frequency of the three peaks as well as their respective amplitudes. Using the same frequencies and widths for the red sidebands, we determine their amplitude, again from a least-squares fit to the data (red curve), which gives $\bar n_\text{\{x,y,z\}}=\{0.04(1),0.02(1),0.06(1)\}$. Assuming a thermal distribution, this means that the atom is cooled to the 1D ground state with a probability of $\{0.96(1),0.98(1),0.95(1)\}$ and to the 3D ground state with a probability of $(89\pm2)\,\%$.

In summary, we have deterministically localized a neutral atom in a 3D optical lattice at the center of a high-finesse optical cavity and have cooled it to the motional ground state of the trapping potential, thus achieving constant and strong single-atom single-photon coupling. Our experiment is the first that simultaneously achieves quantum control over the internal state, position, and momentum of a single atom and over its coupling to light. This is an important step in the development of a truly deterministic light-matter quantum interface \cite{ritter_elementary_2012, nolleke_efficient_2013} with highly improved single-photon absorption and emission efficiencies. Moreover, the exquisite localization of the atom now allows us to realize proposals which require both constant coupling and optical phase stability, such as the generation of entangled states of several atoms in one cavity \cite{pellizzari_decoherence_1995, sorensen_probabilistic_2003, kastoryano_dissipative_2011, nikoghosyan_generation_2012} or the implementation of cavity-based two-qubit quantum gates \cite{pellizzari_decoherence_1995, duan_scalable_2004, duan_robust_2005}.

\begin{acknowledgments}
We acknowledge Manuel Uphoff for his contributions during an early stage of the experiment, Andreas Neuzner for his design of a stable radio-frequency source, and André Kochanke for his design of the microscope objective. This work was supported by the Deutsche Forschungsgemeinschaft (Research Unit 635), by the European Union (Collaborative Projects AQUTE and SIQS), and by the Bundesministerium f\"ur Bildung und Forschung via IKT 2020 (QK\_QuOReP).
\end{acknowledgments}

\end{document}